\documentclass[journal=NanoLetters,manuscript=article]{achemso}
\setkeys{acs}{keywords = true}
\usepackage[version=3]{mhchem} 
\usepackage[T1]{fontenc}       
\usepackage{multirow}
\usepackage{textcomp}

\author{Haoyuan Zhong}
\affiliation
{State Key Laboratory of Low-Dimensional Quantum Physics and Department of Physics, Tsinghua University, Beijing 100084, P. R. China}

\author{Changhua Bao}
\affiliation
{State Key Laboratory of Low-Dimensional Quantum Physics and Department of Physics, Tsinghua University, Beijing 100084, P. R. China}

\author{Huan Wang}
\affiliation
{Department of Physics and Beijing Key Laboratory of Opto-electronic Functional Materials and Micro-nano Devices, Renmin University of China, Beijing 100872, P. R. China.}

\author{Jiaheng Li}
\affiliation
{State Key Laboratory of Low-Dimensional Quantum Physics and Department of Physics, Tsinghua University, Beijing 100084, P. R. China}
\author{Zichen Yin}
\affiliation
{State Key Laboratory of Low-Dimensional Quantum Physics and Department of Physics, Tsinghua University, Beijing 100084, P. R. China}

\author{Yong Xu}
\affiliation
{State Key Laboratory of Low-Dimensional Quantum Physics and Department of Physics, Tsinghua University, Beijing 100084, P. R. China}
\alsoaffiliation
{Frontier Science Center for Quantum Information, Beijing 100084, P. R. China}

\author{Wenhui Duan}
\affiliation
{State Key Laboratory of Low-Dimensional Quantum Physics and Department of Physics, Tsinghua University, Beijing 100084, P. R. China}
\alsoaffiliation
{Frontier Science Center for Quantum Information, Beijing 100084, P. R. China}

\author{Tian-Long Xia}
\affiliation
{Department of Physics and Beijing Key Laboratory of Opto-electronic Functional Materials and Micro-nano Devices, Renmin University of China, Beijing 100872, P. R. China.}

\author{Shuyun Zhou}
\affiliation
{State Key Laboratory of Low-Dimensional Quantum Physics and Department of Physics, Tsinghua University, Beijing 100084, P. R. China}
\alsoaffiliation
{Frontier Science Center for Quantum Information, Beijing 100084, P. R. China}
\email{syzhou@mail.tsinghua.edu.cn}
\title[An \textsf{achemso} demo]
  {Light-tunable surface state and hybridization gap in magnetic topological insulator  MnBi$_8$Te$_{13}$}

\keywords{magnetic topological insulator, MnBi$_8$Te$_{13}$, $\mu$-TrARPES, gap filling, light-tunable interlayer interaction\\}

\begin{document}


\begin{tocentry}
\centering
  \includegraphics[width=8cm]{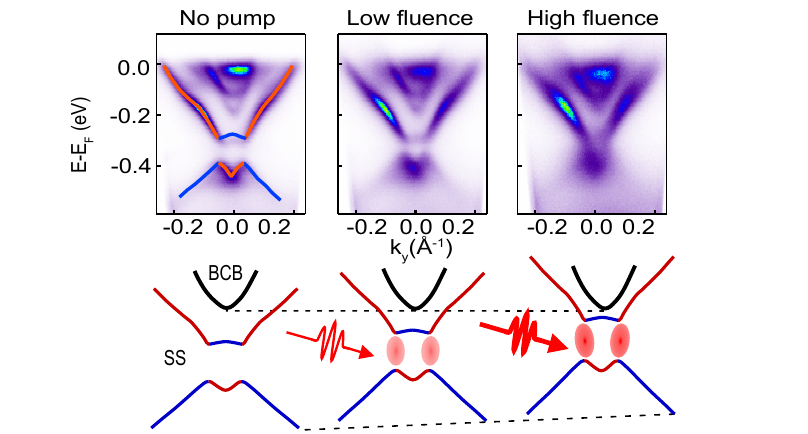}

\end{tocentry}
\label{abstract}
\begin{abstract}
MnBi$_8$Te$_{13}$ is an intrinsic ferromagnetic (FM) topological insulator with different complex surface terminations. Resolving the electronic structures of different termination surfaces and manipulation of the electronic state are important. Here, by using micrometer-spot time- and angle-resolved photoemission spectroscopy ($\mu$-TrARPES), we resolve the electronic structures and reveal the ultrafast dynamics upon photo-excitation. Photo-induced filling of the surface state hybridization gap is observed for the Bi$_2$Te$_3$ quintuple layer directly above MnBi$_2$Te$_4$ accompanied by a nontrivial shift of the surface state, suggesting light-tunable interlayer interaction. Relaxation of photo-excited electrons and holes  is also observed  within 1-2 ps. Our work reveals photo-excitation as a potential control knob for tailoring the interlayer interaction and surface state of MnBi$_8$Te$_{13}$.

\end{abstract}


\label{intro}

Intrinsic magnetic topological insulator (TI) MnBi$_2$Te$_4$ has attracted enormous attention due to the coexistence of topology and magnetism\cite{xue,xu,nature}, which leads to novel quantum phenomena such as quantum anomalous Hall effect (QAHE)\cite{yuanbo} and axion insulator state\cite{yayu}.  Application of an external magnetic field can induce an antiferromagnetic (AFM)  to ferromagnetic (FM) transition, turning MnBi$_2$Te$_4$ from an axion insulator to a Chern insulator with QAHE\cite{yuanbo,yayu,wangjian,helical}. In order to achieve zero-field QAHE, different pathways such as doping\cite{sb,sbweyl} and heterostructures based on magnetic compounds\cite{vbt,cri3} have been proposed to engineer the magnetic state.  Stacking  MnBi$_2$Te$_4$ and  Bi$_2$Te$_3$ to form MnBi$_2$Te$_4$/(Bi$_2$Te$_3$)$_m$ ($m$=1, 2, 3, ...) van der heterostructure is an important pathway to tune the interlayer magnetic interaction.  While MnBi$_4$Te$_7$ (m=1)\cite{dh,147zhaoyue,147nini,147yl,147isava,147jfh,147nini2,147chul} and MnBi$_6$Te$_{10}$ (m=2)\cite{147jfh,147nini2,147chul,1610kaming,1610qiantian} have been found to be AFM similar to MnBi$_2$Te$_4$ (m=0), MnBi$_8$Te$_{13}$ (m=3) clearly distinguishes from other compounds with a unique FM interaction between neighboring MnBi$_2$Te$_4$ layers and QAHE at zero field has been predicted at low temperature\cite{1610nini,1813chaoyu2}, highlighting the importance of interlayer interaction. 

Resolving the electronic structure of MnBi$_8$Te$_{13}$ and manipulation of its band structure in particular the interlayer interaction are both important. MnBi$_8$Te$_{13}$ shows four cleaving surfaces with very different electronic structures\cite{1610nini,1813chaoyu,1813chaoyu2}. Micrometer  ($\mu$m) spot angle-resolved photoemission spectroscopy ($\mu$-ARPES) using laser source is critical for resolving the electronic structures of these different terminations.  Moreover, by adding a pump pulse to  excite the material out of equilibrium and manipulate the transient electronic structure, micrometer spot time- and angle-resolved photoemission spectroscopy ($\mu$-TrARPES)  also allows to reveal the light-matter interaction as well as the relaxation dynamics.  While light-induced spectral weight transfer associated with magnetic ordering has been reported in MnBi$_2$Te$_4$ (m=0)\cite{TR}, so far there is no report on the ultrafast dynamics of MnBi$_8$Te$_{13}$ yet.  Such understanding of the relaxation dynamics upon photo-excitation is important, in particular considering that magnetic topological insulator is a promising candidate for next-generation topological device applications \cite{NRP}. What is particular to MnBi$_8$Te$_{13}$ is that it has four cleaving surfaces with very different electronic structures\cite{1610nini,1813chaoyu,1813chaoyu2}, making it possible to investigate manipulation of the hybridized band structure upon photo-excitation via tuning the interlayer interaction.

Here by using $\mu$-TrARPES, we resolve the electronic structure of MnBi$_8$Te$_{13}$ and reveal the electronic dynamics upon photo-excitation. Interestingly, a filling of the hybridization gap upon photo-excitation is observed for the first Bi$_2$Te$_3$ quintuple layer directly above MnBi$_2$Te$_4$, suggesting light-induced modification of the interlayer interaction. Our work reveals the light-matter interaction in MnBi$_8$Te$_{13}$ and suggests photo-excitation as a useful control knob to manipulate the electronic structure.

\begin{figure*}[htbp]
  \centering
  \includegraphics[]{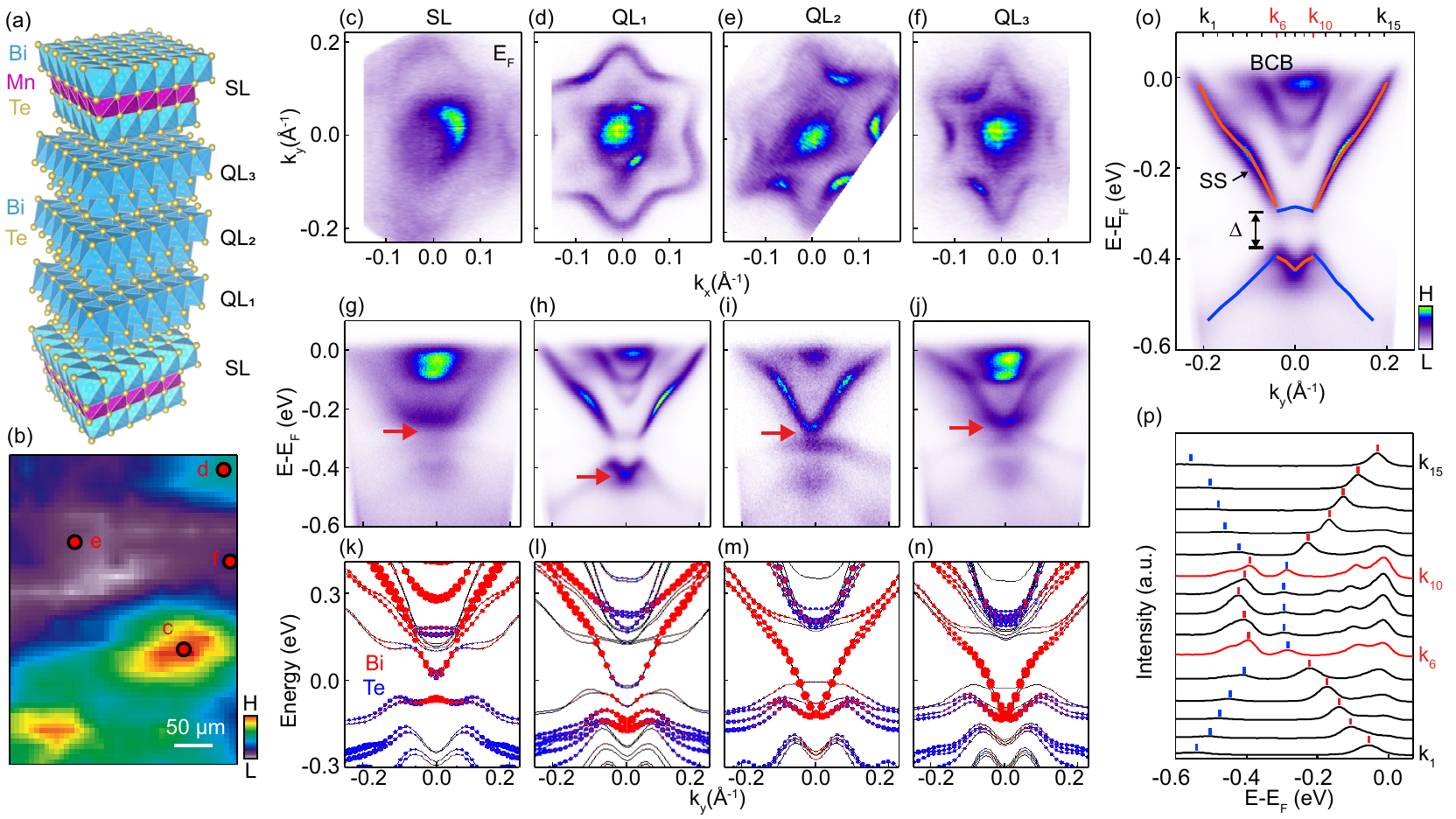}
  \caption{{Spatially-resolved electronic structures of four terminations in MnBi$_8$Te$_{13}$ measured at 80 K.}  {(a)} Crystal structure of MnBi$_8$Te$_{13}$.  {(b)} Spatially-resolved intensity map of MnBi$_8$Te$_{13}$ by integrating ARPES intensity from E$_F$ to -0.1 eV.  {(c-f)} Fermi surface maps of four terminations measured on spots indicated in (b).  {(g-j)} Dispersion images measured along $M$-$\Gamma$-$M$ direction for the four different terminations. The red arrows point to the surface conduction band (SCB) bottom.  {(k-n)} Calculated dispersions along $M$-$\Gamma$-$M$ for the four terminations.  Red and blue circles denote contribution from Bi and Te atoms  on the surface, respectively.  {(o)} Electronic structure of QL$_1$ termination and illustration of its hybridization gap.  {(p)} EDCs  at momentum positions from  k$_1$ to k$_{15}$ indicated in (o).}
\label{fig1}
\end{figure*}

\label{Tex1}

MnBi$_8$Te$_{13}$ single crystals were grown by flux method with a ferromagnetic transition temperature of 10.8 K (see Fig.~S1 for sample characterization). The unit cell of MnBi$_8$Te$_{13}$ consists of one MnBi$_2$Te$_4$ septuple layer (labeled as SL) and three  Bi$_2$Te$_3$ quintuple (QL) layers (labeled as QL$_1$, QL$_2$, QL$_3$ as shown in Fig.~1(a)), which give rise to a mixture of four possible cleaving surfaces.  The electronic structure is not only sensitive to the top termination layer, but also sensitive to the neighboring layer, and therefore all three Bi$_2$Te$_3$ surfaces are expected to exhibit different electronic structures\cite{1610nini,1813chaoyu,1813chaoyu2}. Therefore, resolving the electronic structures for different termination surfaces is the first step before investigating its light-mater interaction. To resolve the different cleaving surface terminations and reveal the corresponding electronic structures, $\mu$-ARPES measurements are performed. Figure 1(b) shows a spatially-resolved intensity map, where different domains with domain size ranging from tens to hundreds of $\mu$m are resolved.  Figure 1(c-f) and Fig.~1(g-j) show the corresponding Fermi surface maps and dispersions measured at four spots as indicated in Fig.~1(b). The Fermi surface map measured in region c (Fig.~1(c)) shows a circular pocket surrounded by a weak flower-shaped pocket, which is consistent with the SL termination\cite{147jfh,147nini,147zhaoyue}.  Similar flower-shaped Fermi pockets are also observed in other domains with QL terminations yet with a decreasing pocket size from QL$_1$ to QL$_3$.  For these three QL terminations, a conical-shaped dispersion from the surface state (SS) is observed in Fig.~1(h-j), with the surface conduction band (SCB) bottom (pointed by red arrow) shifted toward the Fermi energy.  The measured electronic dispersions of these four termination surfaces agree well with the corresponding calculated dispersion for each termination as shown in Fig.~1(k-n), confirming the assignment of the termination surfaces to SL, QL$_1$, QL$_2$ and QL$_3$ respectively. The capability to resolve the electronic structure of each cleaving surface paves the way to further  investigate the effect of light-matter interaction on the electronic structure. 

Among these four termination surfaces, QL$_1$ is particularly interesting with a clearly gapped SS, which is distinguished from other termination surfaces.  The dispersion image  in Fig.~1(o) shows three pockets with parabolic dispersions near the $\Gamma$ point which are from the bulk conduction band (BCB), and there is a conical SS with a much larger pocket size (red and blue curves in Fig.~1(o)) which is the main focus of this work. The SS  shows an overall stronger intensity for the SCB than the surface valence band (SVB), suggesting that they have different orbital contributions. First-principles calculations suggest that the SCB (red curve in Fig. 1(o)) originates from the top QL$_1$ layer while the SVB (blue curve in Fig. 1(o)) originates from the SL underneath\cite{1610nini}. Moving toward the $\Gamma$ point, there is a switching of the intensity contrast  at momentum positions $k_6$ and $k_{10}$, between which the SVB is stronger while  outside this region, the SCB is stronger.   Such intensity inversion suggests that there is a strong hybridization between the SCB and SVB, resulting in a band inversion. A strong suppression of intensity is also observed near the Dirac point,  indicating a gap opening induced by such hybridization. The dispersions and the hybridization gap can be revealed by following the peak positions in the energy distribution curves (EDCs) shown in Fig.~1(p). It is clear that the SCB and SVB have a finite separation, and the size of the gap is extracted to be 105 $\pm$ 5 meV from the peak separation at $k_6$ and $k_{10}$ where the intensity inversion is observed. This gap originates from the hybridization between SL layer and QL$_1$ layer\cite{1610nini}, and a similar gap opening is also observed in the topological SS of the QL$_1$ termination for MnBi$_4$Te$_7$\cite{147chul,147jfh,147zhaoyue} and MnBi$_6$Te$_{10}$\cite{147chul,1610kaming,1610qiantian,1813chaoyu2}. In the following, we reveal the evolution of the gap upon photo-excitation as well as the relaxation dynamics for the BCB and SS in  MnBi$_8$Te$_{13}$. 

\begin{figure*}[htbp]
  \centering
  \includegraphics[]{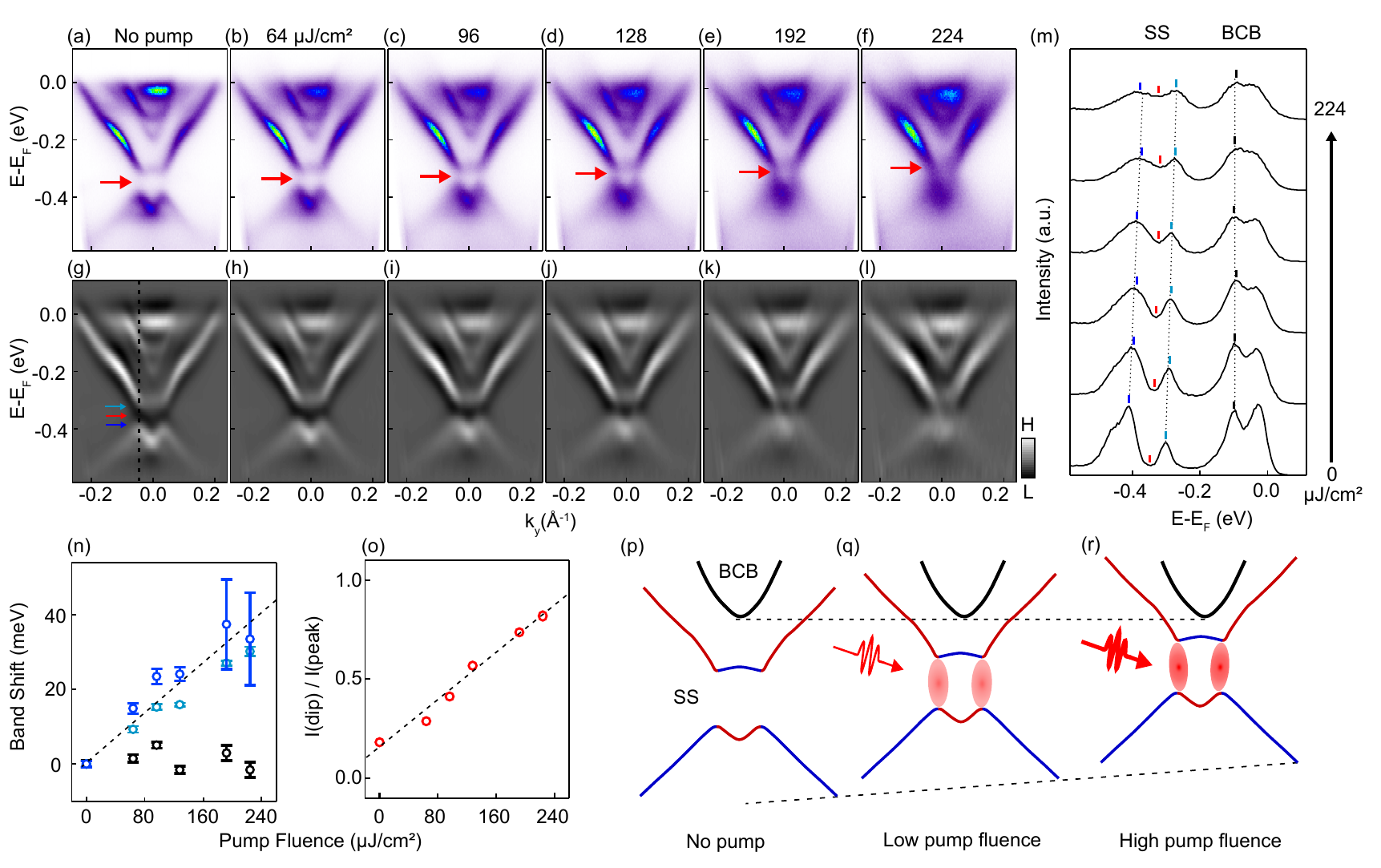}
\caption{{Light-induced gap filling and shift of SS in the QL$_1$ termination.} {(a-f)} Dispersion images of QL$_1$ termination along $M$-$\Gamma$-$M$ measured without a pump pulse (a) and with a pump pulse at increasing pump fluence upon photo-excitation (b-f) measured at zero delay time.  {(g-l)} Corresponding 2D curvature of data in (a-f). {(m)} EDCs extracted from momentum position indicated by black dashed line in (g) at different pump fluence.  Blue tick marks represent peak positions of the SS and red tick marks point to suppression of intensity inside the gap region.  {(n)} Band shift as a function of pump fluence for the SS (blue symbols) and BCB (black symbols). {(o)} Intensity ratio between the dip and the peak as a function of pump fluence. {(p-r)} A schematic illustration of electronic structures observed at different pump fluence. Black curve is the BCB, while red and blue curves are the hybridized surface states.}
\label{fig2}
\end{figure*}

\label{Tex2}

The evolution of the SS for QL$_1$ upon photo-excitation is revealed by $\mu$-TrARPES measurements in Fig.~2, where the largest change is observed near the Dirac point. First of all, the suppression of intensity inside the gap region (indicated by red arrows in Fig.~2(a)) becomes much less obvious in the ARPES dispersion images (Fig.~2(b-f)) and curvature images (Fig.~2(h-l)) with increasing pump fluence, and there is a gradual filling of intensity inside the gap. Eventually at the highest pump fluence of 224 $\mu$J/cm$^2$, the gap becomes almost undetectable. Secondly, both the SCB and SVB shift toward the Fermi energy with pump fluence, while the BCB remains almost the same. The gap filling and shift of the SS are better resolved in the EDCs at the momentum position marked by black dash line in Fig.~2(g), which cuts through the bottom of the SCB and the top of the SVB.  Clear shift in the peak positions of the SCB and SVB can be distinguished in the EDCs shown in Fig.~2(m), while those peaks from the BCB (black dashed line around -0.1 eV in Fig.~2(m)) show negligible change.  The energy shift for both SCB and SVB  increases with pump fluence (light and dark blue dots in Fig.~2(n)) and reaches a large value of 30 meV at the highest pump fluence, while the BCB remains almost unchanged (black dots in Fig.~2(n)). Such nontrivial shift of the SS but not the BCB is in contrast to electron doping of Bi$_2$Se$_3$ where a rigid shift is observed for both the SS and BCB\cite{wangeryin}, and is similar to the surface photovoltaic effect (SPV) reported in Bi$_2$Se$_3$ and Bi$_2$Te$_2$Se\cite{spv1,spv2,spv3,spv4}.  To further quantify the gap filling, we show in Fig.~2(o) the intensity ratio between the dip and the peak, and a linear dependence on the pump fluence is observed. The observed light-induced band modulation of the SS is schematically summarized in Fig.~2(p-r): photo-excitation results in a filling of the gap region between the SCB and SVB, which is also accompanied by a nontrivial shift of the SS toward the Fermi energy while maintaining the energy position of the BCB.

To check if such gap filling could be simply explained by a heating effect induced by the pump pulse, we have extracted the transient electronic temperature in Fig.~S2 from the $\mu$-TrARPES measurements, which shows a electronic temperature of 200 - 280 K.  A comparison of  $\mu$-ARPES measurements at 80 K and 200 K in Fig.~S3 shows that there is negligible change in the hybridization gap, thereby ruling out laser heating as an explanation for the filling of the hybridization gap. Instead, the filling of the hybridization gap suggests that photo-excitation likely modifies the interlayer interaction between Bi$_2$Te$_3$ and the MnBi$_2$Te$_4$ layer underneath. To further check if this speculation is valid, theoretical calculation of the electronic structure with different interlayer spacing has been performed.  The calculated electronic structure shows that an increase of the interlayer spacing between MnBi$_2$Te$_4$ and Bi$_2$Te$_3$ leads to a stronger contribution from the surface Bi$_2$Te$_3$ layer within the hybridization gap (see Fig.~S5 and related discussions in supporting information), supporting that the experimentally observed filling of the hybridization gap is likely caused by an increase of the interlayer spacing. We note that the interlayer spacing  is determined by the van der Waals force between the layers\cite{PRB1979}, which could be modulated by photo-doping as has been demonstrated in MoS$_2$\cite{irmos2} and WSe$_2$\cite{wse2}. In our case, the observed photo-induced energy shift of the surface state in Fig.~2(n) suggests a decrease of carrier concentration (Fig.~S4).  Considering that an increase of carrier concentration leads to a decrease of the interlayer spacing\cite{irmos2}, we would expect that the decrease of carrier concentration in our case leads to an increase of the interlayer distance, resulting in the gap filling as revealed by the calculations (see details in supporting information Fig.~S5).  Therefore, based on the experimental observations and theoretical calculations, we propose light-induced interlayer spacing change as the underlying mechanism for the light-induced filling of the hybridization gap.

\begin{figure*}[htbp]
  \centering
  \includegraphics[]{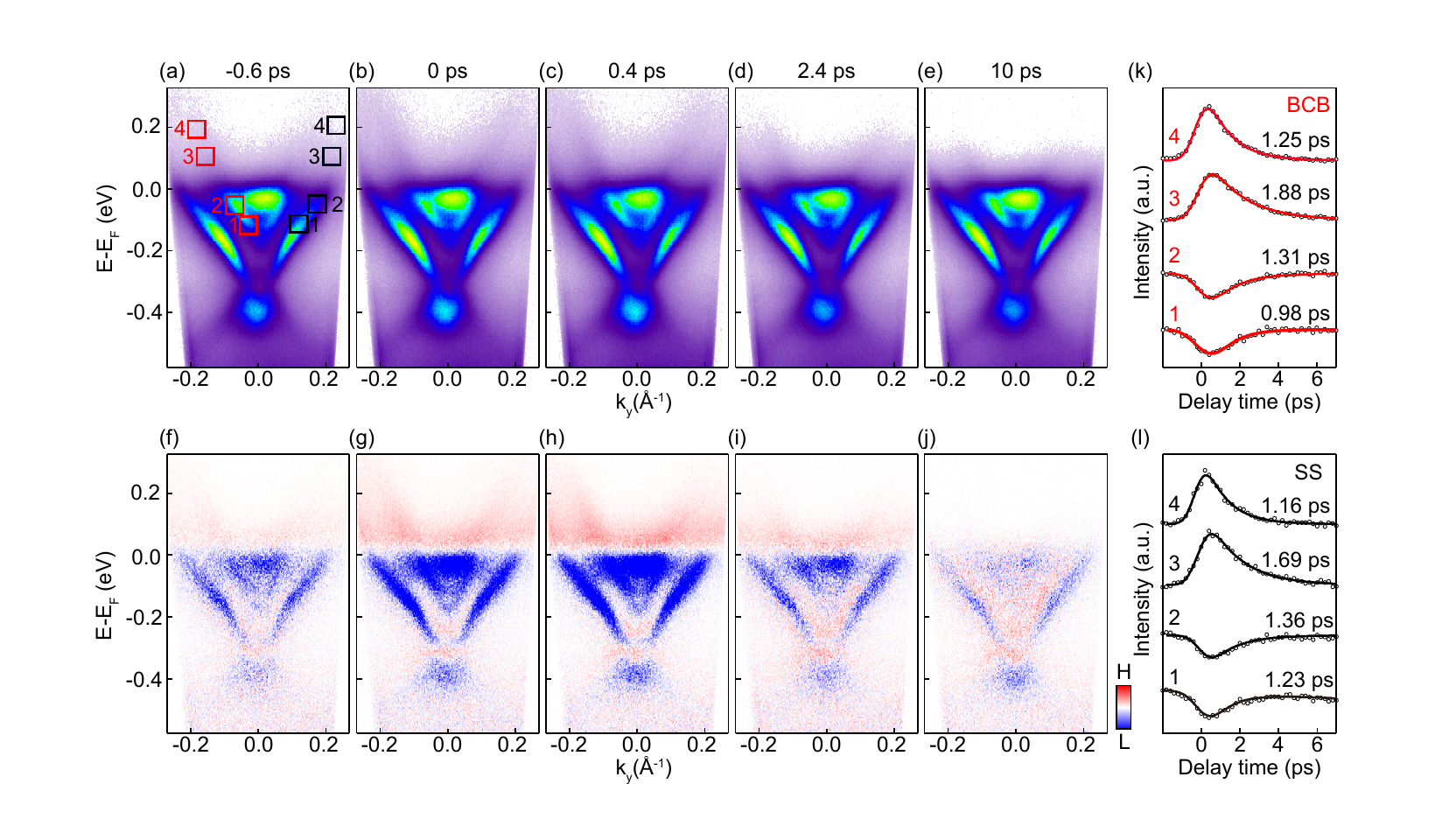}
  \label{Fig3}
\caption{{Ultrafast relaxation dynamics for the BCB and SS upon photo-excitation at pump fluence of 160 $\mu$J/cm$^2$.}  {(a-e)} Snapshots of electronic dispersion in QL$_1$ termination at different pump-probe delay times.  {(f-j)} Differential spectrum image by subtracting the spectrum at a negative delay time of -3.3 ps.  {(k,l)} Temporal dynamics of the electron and hole populations at four different energies obtained by integrating intensity over boxes marked in (a) for BCB (k) and SS (l).}
\end{figure*}

\label{Tex3}
To further reveal the dynamics upon pump excitation, we show in Figure 3 the evolution of the electronic structure with the delay time together with analysis of the relaxation dynamics for photo-excited electrons and holes. Figure 3(a-e) shows transient electronic dispersion images measured along $M$-$\Gamma$-$M$ at different delay times. These data show clearly that the filling of the SS hybridization gap is observed at all delay times, indicating that it is a quasi-static effect with timescale larger than 13 ns, which is the temporal separation between laser pulses. In addition, the excitation and relaxation dynamics of photo-excited carriers, which is important for device applications,  is also revealed by $\mu$-TrARPES measurements. Upon photo-excitation, electrons are excited to the unoccupied states above the Fermi energy, and they relax to lower energy through electron-electron and electron-phonon interaction\cite{BSzx,BSgedik,BShofmann}, and a much longer lifetime is in general expected for photo-excited states near the Fermi energy.  The photo-excited electronic states can be better revealed in the differential images in Fig.~3(f-j), which are obtained by subtracting the dispersion image by data measured at -3.3 ps. Photo-excited holes (blue) and electrons (red) are clearly resolved and the excitation is  not limited to states around the Fermi energy but rather it involves an extended energy range of the SS as deep as -0.4 eV. Carrier dynamics of different bands are directly visualized by tracking the continuous time evolution of photo-excited electrons and photo-excited holes for the BCB (Fig.~3(k)) and the SS (Fig.~3(l)) by integrating over selected energy and momentum regions marked by red and black boxes in Fig.~3(a). The time trace is fitted by a Gaussian function convolved with the product of the step function and an exponential function $
I(t)=A(1+erf(\frac{t-t_0}{\Delta t}-\frac{\Delta t}{2\tau}))e^{-\frac{t-t_0}{\tau}}+B$ \cite{timeresolution}, 
 where $\Delta t$ is the width of the rising edge and $\tau$ is the relaxation time. The relaxation time of the photo-excited electrons in the BCB increases from 1.25 to 1.88 ps when moving towards the Fermi energy and increases from 0.98 to 1.31 ps for photo-excited holes, which is similar to the relaxation time ranging from 1.16 to 1.69 ps for for photo-excited electrons and 1.23 to 1.31 ps for photo-excited holes of the SS. 
  The relaxation time scale of picoseconds is similar to that observed in MnBi$_2$Te$_4$\cite{TR} and topological insulators such as Bi$_2$Se$_3$\cite{BSzx,BSgedik,BShofmann} and Bi$_2$Te$_2$Se\cite{BTS1,BTS2} and suggests a relaxation through both electron-electron and electron-phonon coupling.  In addition, the revealed relaxation dynamics with an ultrafast fast relaxation lifetime of a few ps provides key parameters for future device applications\cite{NRP}.

\label{summary}
In summary, by using $\mu$-TrAPRES, we reveal the electronic structures of different termination surfaces for  MnBi$_8$Te$_{13}$ and report light-induced filling of the hybridization gap in the QL$_1$ surface, which is also accompanied by a nontrivial shift of the SS while the BCB remains unchanged.  The light-induced filling of the hybridization gap is likely caused by the light-induced modification of the interlayer coupling strength, which is supported by calculated electronic structure at different interlayer spacings. We note that light-induced modification of interlayer interaction has been reported in ZrTe$_5$ though photon-phonon coupling\cite{ztprx}, in MoS$_2$ by changing the polarizability\cite{irmos2} and in WTe$_2$ by an interlayer shear strain\cite{LindenbergMoTe2}, and light-tunable interlayer interaction has also been proposed for engineering the electronic structure of magic-angle twisted bilayer graphene\cite{Fiete2020PRB,Fiete2020PRR}. 
Our work adds a new experimental example of manipulating the electronic structure by tuning interlayer interaction using photo-excitation in a magnetic topological insulator which provides new opportunities for exploring possible photo-induced ultrafast magnetic modulation\cite{mbtkerr} and topological transition.



\begin{suppinfo}
The Supporting Information is available free of charge at http://pubs.acs.org.

Sample characterization including X-Ray diffraction pattern, temperature dependent magnetic susceptibility and magnetic hysteresis loop, electronic temperature in the TrARPES measurements and temperature-dependent ARPES measurements, possible mechanism of light-induced gap filling and time resolution of TrARPES system.
\end{suppinfo}

\begin{acknowledgement}
This work is supported by the National Key R $\&$ D Program of China (Grants 2016YFA0301004, 2016YFA0301001, 2020YFA0308800), the National Natural Science Foundation of China (Grants 11725418 and 11427903),  Beijing Advanced Innovation Center for Future Chip (ICFC) and Tsinghua University Initiative Scientific Research Program, and Tohoku-Tsinghua Collaborative Research Fund. T.-L.X. is supported by the National Natural Science Foundation of China (Grant 12074425) and  the National Key R $\&$ D Program of China (Grant 2019YFA0308602). W.D. is supported by the National Natural Science Foundation of China (Grant 51788104).
\end{acknowledgement}

\section{METHODS}

\subsection{ARPES measurement}
\quad $\mu$-ARPES and $\mu$-TrARPES measurements were performed in our home laboratory at Tsinghua University using a Ti-sapphire oscillator producing femtosecond pulses at 745 nm (1.66 eV) at 76 MHz repetition rate. The infrared laser was frequency quadrupled to produce ultraviolet probe laser by nonlinear optical crystal BBO and KBBF with photon energy $h\nu$ = 6.66 eV. The pump and probe beams are both p-polarized and are focused onto the sample with a beam size of 30 $\mu$m $\times$ 40 $\mu$m and 10 $\mu$m $\times$ 20 $\mu$m (full width half maximum) respectively. The overall temporal resolution is 611 fs determined by Sb$_2$Te$_3$ thin film (see in Fig.~S6) and the energy resolution is 45 meV. The tunable pump-probe delay was achieved by varying the pump optical path with a motorized delay stage of 6.7 fs precision. ARPES measurements were performed on freshly cleaved MnBi$_8$Te$_{13}$ single crystals at 80 K with a base pressure better than $5\times 10^{-11}$ Torr. 

\subsection{Sample growth}
\quad The MnBi$_8$Te$_{13}$ single crystals were grown by flux method. Mn powder, Bi lump and Te lump were weighed with the ratio of MnTe:Bi$_2$Te$_{3}$=19:81. The mixtures were loaded into a corundum crucible which was sealed into a quartz tube. Then the tube was put into a furnace and heated to 1100 \textcelsius~ for 20 hours to allow sufficient homogenization. After a quick cooling to 600 \textcelsius~ at 5 \textcelsius/hour, the mixtures were slowly cooled down to 578 \textcelsius\ at 0.5 \textcelsius/hour and kept for 2 days. Finally, the single crystals were obtained after centrifuging. The plate-like MnBi$_8$Te$_{13}$ single crystals with centimeter scale are characterized by single crystal X-ray diffraction (XRD), powder XRD together and magnetic measurements as shown in Fig.~S1.

\subsection{First-principles calculations}
\quad First-principles calculations were performed in the framework of density functional theory using the Vienna ab initio Simulation Package (VASP)\cite{1}. The plane-wave basis with the energy cutoff of 350 eV was adopted. The projector augmented wave potential was used to simulate the ion core environment, accompanied by the Perdew-Burke-Ernzerhof exchange-correlation functional. The localized 3d-orbitals on the Mn atoms were described by the GGA+$U$ methods\cite{2} and the $U$ value was selected at 3.0 eV. Lattice constants were fixed at the experimental value, and atoms inside were fully relaxed with a force criterion of 0.01 eV/\AA. The DFT-D3 method\cite{3} was introduced to tackle Van der Waals interactions, and spin-orbit-coupling effect was included self-consistently. The Monkhorst-Pack k-point mesh of 11$\times$11$\times$3 was adopted for the bulk structure and 11$\times$11$\times$1 for films. Surface state calculations were performed by the WannierTools package\cite{4}, based on the tight-binding Hamiltonians constructed from maximally localized Wannier functions.

%
%
%


\providecommand{\latin}[1]{#1}
\makeatletter
\providecommand{\doi}
  {\begingroup\let\do\@makeother\dospecials
  \catcode`\{=1 \catcode`\}=2 \doi@aux}
\providecommand{\doi@aux}[1]{\endgroup\texttt{#1}}
\makeatother
\providecommand*\mcitethebibliography{\thebibliography}
\csname @ifundefined\endcsname{endmcitethebibliography}
  {\let\endmcitethebibliography\endthebibliography}{}

\end{document}